# The standard genetic code facilitates exploration of the space of functional nucleotide sequences

Shubham Tripathi[1, 2] and Michael W. Deem[2, 3, 4, *]

[1] PhD Program in Systems, Synthetic, and Physical Biology, Rice University, Houston, TX 77005, USA.

[2] Center for Theoretical Biological Physics, Rice University, Houston, TX 77005, USA.

[3] Department of Bioengineering, Rice University, Houston, TX 77005, USA.

[4] Department of Physics and Astronomy, Rice University, Houston, TX 77005, USA.

[*] Corresponding author. Email: mwdeem@rice.edu

# Abstract

The standard genetic code is well known to be optimized for minimizing the phenotypic effects of single nucleotide substitutions, a property that was likely selected for during the emergence of a universal code. Given the fitness advantage afforded by high standing genetic diversity in a population in a dynamic environment, it is possible that selection to explore a large fraction of the space of functional proteins also occurred. To determine whether selection for such a property played a role during the emergence of the nearly universal genetic code, we investigated the number of functional variants of the *Escherichia coli* PhoQ protein explored at different time scales under translation using different genetic codes. We found that the standard genetic code is highly optimal for exploring a large fraction of the space of functional PhoQ variants at intermediate time scales as compared to random codes. Environmental changes, in response to which genetic diversity in a population provides a fitness advantage, are likely to have occurred at these intermediate time scales. Our results indicate that the ability of the standard code to explore a large fraction of the space of functional sequence variants arises from a balance between robustness and flexibility and is largely independent of the property of the standard code to minimize the phenotypic effects of mutations. We propose that selection to explore a large fraction of the functional sequence space while minimizing the phenotypic effects of mutations contributed towards the emergence of the standard code as the universal genetic code.

# Introduction

The genetic code enumerates the rules for the translation of messenger RNAs into proteins, the mapping from each of the 61 three-nucleotide codons to one of the 20 amino acids (Nirenberg et al. 1963; Woese 1967). The remaining 3 codons signal the end of the protein coding region. The genetic code is one of the universal features of life, with only minor variations across the three domains (Knight et al. 2001; Alberts et al. 2008). The assignment of amino acids to different codons is not random. Even as the genetic code was being deciphered in the 1960s, it was recognized that codons differing by a single base are either assigned the same amino acid or amino acids that are biochemically similar in the standard genetic code, as compared to random genetic codes (Woese 1965; Epstein 1966; Goldberg and Wittes 1966; Alff-Steinberger 1969). Recent studies utilizing computer simulations have lent quantitative support to this notion (Haig and Hurst 1991; Freeland and Hurst 1998; Butler et al. 2009). The organization of codon-amino acid assignments in the standard genetic code may have evolved to minimize, on average, the phenotypic effect of genetic mutations, transcription errors, and mistranslations (Cullmann and Labouygues 1983). Others have suggested that the genetic code co-evolved with pathways for amino acid synthesis, with amino acids having closer precursor-product relationships in biosynthetic pathways being coded by similar codons (Wong 1975; Taylor and Coates 1989; Freeland et al. 2000). Another view is that the codon-amino acid assignment is an outcome of the physiochemical affinity between amino acids and cognate codons (Pelc 1965; Dunnill 1966; Pelc and Welton 1966). Experimental and statistical evidence in favor of this last theory has, however, been inconclusive (Ellington et al. 2000).

There are two prevalent theories of the mechanism via which a universal genetic code may have evolved (Koonin and Novozhilov 2009). The first theory suggests that the universality of the genetic code is an outcome of the fact that all present-day life forms evolved from a universal common ancestor. After the emergence of translation, the genetic code fixed in the population of a single niche and froze. It was only after the code froze that life forms diversified from the single niche initially occupied (Crick 1968; Wong 1976; Harris et al. 2003). This frozen code was then inherited unchanged during the subsequent spread and diversification of life forms. Another possibility in this vertical descent model of genetic code evolution is that the genetic code froze after life had diversified. As life forms spread, organisms with distinctive characteristics emerged, some even lacking translation. Those organisms with translation machineries would have evolved different genetic codes. However, over an extended period of time, all codes except the standard genetic code were lost, either due to neutral drift or due to selection for optimality properties of the standard genetic code.

In contrast to the vertical descent model described above is the view that a universal and optimal genetic code emerged from communal evolution amidst extensive horizontal gene transfer during the early stages of life (Vetsigian et al. 2006; Goldenfeld and Woese 2007). With organisms utilizing different genetic codes in multiple communities competing for a single niche, the community wherein members utilize the same genetic code or compatible genetic codes is more likely to succeed. This is because a community-wide code or a set of compatible codes will allow for efficient sharing of newly evolved beneficial proteins among different individuals in the community via horizontal gene transfer, thereby allowing the organisms in the community access to a larger innovation pool. Further, robustness to mistranslations is likely to be selected for in such a community-wide code due to the likely inefficient translation machineries in these ancient life forms. Simulations have suggested that communal evolution amidst extensive horizontal gene transfer allows for the evolution of a code that is more optimized for minimizing the phenotypic effect of translation errors. In the absence of horizontal gene transfer, conversely, evolving genetic codes tend to get stuck in local minima, ending up less optimal for minimizing the phenotypic effect of translation errors (Vetsigian et al. 2006).

In both mechanisms of emergence of a universal code described above, selection acts on the phenotypic features of the organism, and not directly on the system of codon-amino acid assignments. Mutations in the genomic DNA sequence, or the RNA sequence in the case of some viruses, are inherited by the progeny during replication. Since proteins are the molecules that carry out a majority of the cellular functions, it has been argued that they largely determine the phenotype of the organism (Griffiths et al. 2000). Given that the genetic code governs the translation of the transcript of the genetic material into proteins, it plays a fundamental role in steering molecular evolution (Gonnet et al. 1992). Here, we probe this dependence of molecular evolution on the genetic code.

Life forms have evolved over time amidst changing environmental conditions. Different environmental conditions require different phenotypic responses for an organism to survive. Considering the large, yet finite, space of possible amino acid sequences, a species that can access a greater portion of the sequence space is more likely to encounter a protein capable of forging a fitting response to a new environmental condition. However, only those proteins encountered during the exploration of the sequence space that are functional can contribute towards the survival of the organism. Thus, paths through the sequence space should be highly biased towards those containing functional protein variants. Given the evolutionary advantage of exploring a larger fraction of the space of protein sequences, and the role of the genetic code in guiding molecular evolution, the standard genetic code may have evolved under a selection pressure to maximize the fraction of the functional sequence space explored. We test this hypothesis using the landscape of functional variants of the PhoQ protein.

Previous studies have suggested that the organization of the standard genetic code constrains the exploration of the space of possible amino acid sequences (Maynard Smith 1970). This is because a single nucleotide change allows access to only 6 of the 19 possible amino acid substitutions on average and silent mutations are abundant within the standard genetic code. Maeshiro and Kimura suggested that the standard genetic code allows for a balance of robustness and changeability via a balance between the probabilities of synonymous and non-synonymous changes in amino acids under single nucleotide substitutions (Maeshiro and Kimura 1998). Judson and Haydon found that codes computationally evolved under selection for characteristics such as higher amino acid connectedness and shorter path length between different amino acids were closer to the standard genetic code (Judson and Haydon 1999). Few studies have explored the dependence of the abundance of adaptive mutations on the genetic code. Zhu and Freeland, using a population genetic code model, found that the standard genetic code has properties that enhance the efficacy of adaptive sequence evolution (Zhu and Freeland 2006). Firnberg and Ostermeier showed that the standard genetic code enriches for adaptive mutations in the antibiotic resistance gene TEM-1 $\beta$-lactamase (Firnberg and Ostermeier 2013).

The aim of the present study was to characterize how the genetic code affects the exploration of the space of functional amino acid sequences. We considered the exploration of functional variants of the *Escherichia coli* protein kinase PhoQ under translation using different genetic codes. The standard genetic code explored a larger fraction of the functional protein sequence space at small and intermediate time scales, compared to random genetic codes with the same degeneracy as the standard code and to random genetic codes with degeneracies different from that of the standard code. Upon considering longer time scales, the fraction of random codes of both types that allowed for the exploration of more functional PhoQ variants than the standard code increased. However, less than 5% of the random genetic codes with the same degeneracy as the standard code allowed for the exploration of more functional PhoQ variants than the standard code even at these extended time scales. We also investigated the dependence of the fraction of the functional sequence space explored on the starting nucleotide sequence. Finally, we calculated the correlations of the fraction of the sequence space explored under translation using different genetic codes with different quantitative characteristic features of the genetic code.

# Results

We generated random genetic codes using two different approaches, one that preserved the degeneracy of the standard genetic code ($T_{DP}$ codes) and one that did not preserve the degeneracy ($T_{DNP}$ codes). The stop codons were left unchanged in both the approaches. Podgornaia and Laub probed the PhoP binding ability of all 160,000 possible variants of the PhoQ protein with amino acid substitutions at the positions 284, 285, 288, and 289 (Podgornaia and Laub 2015). Only 1659 variants of PhoQ with substitutions at these positions were functional. Starting from a nucleotide sequence coding for the original PhoQ amino acid sequence in *E. coli*, hereafter referred to as the wild type PhoQ variant, we introduced single nucleotide substitutions in the 12-nucleotide long sequence. Since multi-nucleotide substitutions are rare

(Terekhanova et al. 2013), we do not consider them here. After each mutation, the nucleotide sequence was translated into the corresponding amino acid sequence using the code under consideration. If the new protein was functional, the mutated nucleotide sequence became the starting sequence in the next simulation step. Otherwise, the non-mutated nucleotide sequence was carried forward to the next step. Simulations were carried out for 100, 1000, 10000, 100000, and 1000000 steps.

**The standard genetic allows for the exploration of more functional variants of PhoQ as compared to random codes.**

Fig. 1 shows the distribution of the number of functional variants of the PhoQ protein explored under translation using different random genetic codes and using the standard genetic code for different numbers of simulation steps. Under translation using the standard genetic code, more unique functional variants of PhoQ were explored via single nucleotide substitutions as compared to the average number visited under translation using randomly generated codes of both types, $T_{DP}$ and $T_{DNP}$, for up to 100000 simulation steps. It is only at 1 million simulation steps the the average number of functional PhoQ variants visited under translation using 10000 type $T_{DNP}$ codes surpassed the number explored under translation using the standard genetic code. The mean number for type $T_{DP}$ codes, however, remained small as compared to the standard code even at 1 million simulation steps, fig 2 (A). Among codes of type $T_{DP}$, less than 5% allowed the exploration of more functional variants than the standard code for up to 1 million simulation steps. The fraction was lower at lower numbers of simulation steps as is evident from fig. 2 (B). Among codes of type $T_{DNP}$, less than 7% permitted exploration of more functional PhoQ variants that the standard code with number of simulation steps up to 10000. The fraction was larger for higher numbers of simulation steps, reaching 87.44% at 1 million simulation steps. The results in fig. 2 (B) indicate that the standard genetic code is more optimized for exploring a large fraction of the space of functional PhoQ variants at intermediate time scales than at very short or long time scales. The fraction of random codes of both types that explored a larger fraction of the PhoQ functional sequence space as compared to the standard genetic code was lower at 1000 and 100000 simulation steps than at 100, 100000, or 1 million simulation steps, fig 2 (B).

**Number of functional PhoQ variants explored varies only slightly for different degenerate starting nucleotide sequences.**

Since there are only 20 amino acids and 3 stop codons with 64 possible three-nucleotide codons, all genetic codes are degenerate, i.e. an amino acid may be encoded by more than one codon. Therefore, the same amino acid sequence can be encoded by multiple distinct nucleotide sequences. We investigated how the number of functional PhoQ variants explored depends on the nucleotide sequence coding for the wild type PhoQ variant from which the simulation is started. The results are shown in fig. 3. For the standard genetic code, the standard deviation of the number of proteins explored over all 384 possible starting nucleotide sequences was only 3% and 2% of the mean for 100 and 1000 simulation steps respectively. The average standard deviation as a fraction of the mean was 4.57% and 3.75% for type $T_{DP}$ codes, and 12.31% and 12.08% for type $T_{DNP}$ codes, for 100 and 1000 simulation steps, respectively. The distribution of standard deviations as fractions of means was wider for type $T_{DNP}$ codes as compared to type $T_{DP}$ codes.

**The space of functional nucleotide sequences is modular for all genetic codes.**

We next investigated the structure of the space of functional nucleotide sequences under different genetic codes. For each genetic code, we constructed a network with nodes as the nucleotide sequences functional under the given code, and edges between sequences differing by one nucleotide substitution. We calculated the Newman modularity (Newman 2004; Newman and Girvan 2004) of this network using the Louvain algorithm (Blondel et al. 2008) for the standard code, for 100 type $T_{DP}$ codes, and for 100 type $T_{DNP}$ codes. For this calculation, we used the computer code available from http://www.ludowaltman.nl/slm/. As shown in fig. 4, the modularity of the space of functional nucleotide sequences is very high for the standard genetic code and for the randomly generated codes (> 0.85 in all cases). Further, we observed a larger variation in modularity values for type $T_{DNP}$ codes as compared to type $T_{DP}$ codes.

**Codes that better preserve the chemical properties of amino acids under point mutations explore more functional PhoQ variants.**

To determine the factors with which the numbers of functional PhoQ variants explored are correlated, we calculated different characteristic measures for the standard genetic code and for the random genetic codes. These measures can be classified into two categories, ones that characterize the effects of single nucleotide substitutions on physio-chemical properties of the amino acids encoded, and ones that quantify different structural features of codon-amino acid assignments in different genetic codes. We discuss measures of the first type here. For each genetic code, we calculated the average over all single nucleotide substitutions in all codons, except the stop codons, of the squared change in different physical and chemical properties of the amino acids encoded: polar requirement (C R Woese et al. 1966), hydrophilicity (Weber and Lacey 1978), isoelectric point (Alff-Steinberger 1969), and amino acid volume (Zamyatnin 1972).

We found weak, yet statistically significant (p-value < $10^{-4}$) negative correlations between the number of functional PhoQ variants explored and the mean squared change in polar requirement, hydrophilicity, and isoelectric point for both type $T_{DP}$ and type $T_{DNP}$ codes, at all numbers of simulation steps, fig. 5 and fig. 6. The correlation between the number of functional PhoQ variants explored and the mean squared change in amino acid volume was further weak, with statistical insignificance at certain numbers of simulation steps, fig. 5 and fig. 6.

To further probe the relation between the mean squared change in polar requirement for genetic codes and the number of functional PhoQ variants explored, we compared the mean squared changes in polar requirement for the 100 type $T_{DP}$ codes and the 100 type $T_{DNP}$ codes that allowed for the exploration of most PhoQ variants to the mean squared changes in polar requirement for all type $T_{DP}$ and type $T_{DNP}$ codes, fig. 7 (A) and (B). We observed that the mean squared changes in polar requirement for the top performing type $T_{DP}$ and type $T_{DNP}$ codes did not differ significantly from other random codes. We also considered the 100 type $T_{DP}$ and the 100 type $T_{DNP}$ codes with least values of mean squared changes in polar requirement and found that the numbers of functional PhoQ variants explored under translation using these codes did not differ significantly from the numbers explored under translation using all random codes of each type, fig. 7 (C) and (D).

**Codes with a larger number of non-synonymous point mutations explore more functional PhoQ variants.**

We evaluated quantitative measures of different structural features of codon-amino acid assignments in randomly generated genetic codes and investigated their correlation with the number of functional PhoQ variants explored under translation using the different codes. We used the measures defined previously by Judson and Haydon (Judson and Haydon 1999): code fragility, defined as the number of codons with eight or nine non-synonymous point mutations out of the nine possible point mutations; code mutability, defined as the average number of non-synonymous point mutations per codon; and the total number of synonymous point mutations. Scatter plots of these measures versus the number of functional PhoQ variants explored under translation using type $T_{DNP}$ codes at different numbers of simulation steps are shown in fig. 7. Note that all type $T_{DP}$ have the same value as the standard genetic code for the measures considered in this section.

We observed that code fragility and code mutability were positively correlated with the number of functional PhoQ variants visited for different numbers of simulation steps, with p-value $< 10^{-4}$ in each case. The total number of synonymous mutations, which encodes information opposite to that encoded by code fragility and code mutability, was negatively correlated, p-value $< 10^{-4}$, with the number of functional PhoQ variants explored under translation using type $T_{DNP}$ codes for different numbers of simulation steps.

We also calculated the code changeability (Maeshiro and Kimura 1998) for different type $T_{DNP}$ codes. Changeability is defined as the sum, over all pairs, of the probabilities of transitions between different amino acids, excluding paths involving stop codons. The number of functional PhoQ variants explored under translation using type $T_{DNP}$ codes was negatively correlated with code changeability, p-value $< 10^{-4}$, at numbers of simulation steps greater than 100.

**Preservation of chemical properties of amino acids under point mutations facilitates exploration of the functional PhoQ sequence space at intermediate time scales.**

We calculated, for each number of simulation steps, the mean squared change in amino acid polar requirement, hydrophilicity, isoelectric point, and volume due to point mutations under translation using those type $T_{DP}$ and type $T_{DNP}$ codes that explored more functional PhoQ variants as compared to the standard genetic code at the given number of simulation steps. The results are shown in fig. 8. The mean squared change in these physio-chemical properties of amino acids due to point mutations is lower for codes that explored more functional PhoQ variants than the standard genetic code for 1000 and 10000 simulation steps as compared to the codes that explored more functional PhoQ variants than the standard genetic code for 100, 100000, or 1000000 simulation steps. These results indicate that preservation of chemical properties of amino acids under point mutations promotes the exploration of functional PhoQ variants at intermediate time scales while contributing little towards exploration of functional PhoQ variants at very short time scales i.e. 100 simulation steps, or at very long time scales i.e. 100000 and 1000000 simulation steps.

**The deviant genetic codes in different species explore more functional PhoQ variants as compared to the standard genetic code.**

Maeshiro and Kimura (Maeshiro and Kimura 1998) proposed that the reassignment of certain codons in the deviant genetic codes of some species such as *Candida* spp., *Mycoplasma* spp., *Euplotes* spp., and *Blepharisma* spp. could ease the transitions between amino acids with different polarities, and increase chances of recovery from nonsense mutations by decreasing the number of stop codons, thereby allowing for greater alterability of the phenotypes. Postulating that these properties will facilitate the exploration of the space of functional PhoQ variants, we calculated the number of functional PhoQ variants explored under translation using these deviant genetic codes at different numbers of simulations steps and compared the results to the number of functional PhoQ variants explored under translation using the standard genetic code. The results are shown in fig. 8 and in table 1. For numbers of simulation steps greater than 1000, translation using deviant genetic codes in *Candida* spp., *Euplotes* spp., and *Blepharisma* spp. allowed for the exploration of a significantly higher (two-sample t-test p-value < 0.01) fraction of the space of functional PhoQ variants as compared to translation using the standard genetic code. Differences from the standard genetic code are not significant at 100 and 1000 simulation steps for the deviant codes in these species. Under translation using the deviant genetic code in *Mycoplasma* spp., a significantly higher fraction of the PhoQ sequence space as compared to translation using the standard code (two-sample t-test p-value < 0.05) is explored only at 100000 and 1000000 simulation steps. For each code and each number of simulation steps, 100 simulations were carried out and the results used to calculate the p-values.

# Discussion

Our results indicate that the organization of codon-amino acid assignments in the standard genetic code allows for the exploration of a greater fraction of the space of functional variants of the *Escherichia coli* protein kinase PhoQ via single nucleotide substitutions, as compared to most randomly generated codes, particularly at intermediate time scales. A role of selection for flexibility in the evolution of the standard genetic code was first proposed by Maeshiro and Kimura (Maeshiro and Kimura 1998) and backed up soon thereafter by Judson and Haydon (Judson and Haydon 1999). Firnberg and Ostermeier showed for a subset of the functional variants of the antibiotic resistance gene TEM-1 $\beta$-lactamase that there is enrichment for adaptive mutations under translation using the standard genetic code (Firnberg and Ostermeier 2013; Firnberg et al. 2014). Our results represent the first direct confirmation of Maeshiro and Kimura's hypothesis for the space of all possible variants of an amino acid sequence.

Both with randomly generated codes and the standard code, the space of functional nucleotide sequences is partitioned into clusters of sequences with dense connections between nucleotide sequences in the same cluster and sparse connections between sequences in different clusters. This modular structure of the functional nucleotide sequence space for all genetic codes arises since all genetic codes map 61 codons to 20 amino acids and must therefore be degenerate. Given the small separation between degenerate codons in the standard code and in type $T_{DP}$ codes, the probability that a single nucleotide substitution will change the encoded amino acid is low. Further, amino acid sequences that are closer to a functional sequence are more likely to be functional as compared to the sequences that are far away from it. This is since protein functionality derives from the physio-chemical properties of amino acids. Thus, substitution of an amino acid with another amino acid having similar properties is less likely to alter the functionality of the amino acid sequence. This characteristic results in a modular structure of the functional nucleotide sequence space, even for type $T_{DNP}$ codes where degenerate codons may be separated by a larger distance.

The modular structure of the functional nucleotide sequence space is responsible for the comparatively small variation in the number of functional PhoQ variants visited on starting the simulation from different nucleotide sequences coding for the wild type PhoQ amino acid sequence. For the standard code and for type $T_{DP}$ codes, all nucleotide sequences coding for the wild type PhoQ are likely to lie within the same cluster, given the small distances between degenerate codons in these codes. Degenerate codons in these codes differ by 1.3 nucleotides on average. Since our simulation is a random walk on the network of functional nucleotide sequences, given the highly modular nature of the network, simulations starting at nodes within the same cluster are likely to explore similar numbers of nodes. For type $T_{DNP}$ codes, degenerate codons differ by 2.25 ± 0.07 nucleotides (mean ± standard deviation). Thus, different nucleotide sequences coding for the wild type PhoQ are less likely to be located within the same cluster, resulting in a larger variation and a wider distribution of variations in the number of functional PhoQ sequences visited as compared to type $T_{DP}$ codes on starting the simulation from different degenerate nucleotide sequences.

The property of the standard genetic code to allow for the exploration of a large fraction of the space of functional nucleotide sequence variants is not a direct consequence of the property of the code to minimize changes in different physio-chemical properties of the encoded amino acids under point mutations, a property that has been demonstrated in previous studies (Alff-Steinberger 1969; Zamyatnin 1972; Wolfenden et al. 1979; Haig and Hurst 1991; Freeland and Hurst 1998). The fact that the former property does not directly lead to the latter is evident from the weaker optimization of the standard genetic code for exploration of the space of functional PhoQ variants as compared to its optimization for minimizing the changes in physio-chemical properties of amino acids under point mutations. None of the 10000 type $T_{DP}$ or type $T_{DNP}$ codes exhibited a smaller mean squared change in amino acid polar requirement under single nucleotide substitutions than the standard code. The 100 type $T_{DP}$ codes and the 100 type $T_{DNP}$ codes that allowed for the exploration of highest numbers of functional PhoQ variants did not exhibit significantly lower values of the mean squared change in polar requirement as compared all random codes. The 100 type $T_{DP}$ codes and the 100 type $T_{DNP}$ codes with least values of mean squared changes in polar requirement did not allow for the exploration of significantly higher numbers of functional PhoQ variants as compared to all random codes. Further, while all type $T_{DP}$ codes allowed for the same number of synonymous mutations as the standard code, only a few of these codes allowed for the exploration of a larger fraction of the space of functional PhoQ variants than the standard code. None of the type $T_{DNP}$ codes allowed for more synonymous mutations than the standard code, but a greater number of such codes allowed for the exploration of a greater fraction of the space of functional PhoQ variants as compared to the standard code. Taken together, these observations indicate that selection for exploring a large fraction of the space of functional nucleotide sequences is largely independent of the different selection pressures postulated before. In fact, selection for a code that allows for the exploration of a larger fraction of the space of functional protein variants as compared to the standard genetic code may have contributed towards the emergence of codon re-assignments in deviant genetic codes seen in *Candida* spp., *Mycoplasma* spp., *Euplotes* spp., and *Blepharisma* spp.

The correlations of the number of functional PhoQ variants explored with different characteristic measures of the genetic codes indicate that a balance between robustness and flexibility is needed for better exploration of the space of functional nucleotide sequences (Maeshiro and Kimura 1998). While genetic codes that are conservative for changes in polar requirement, hydrophilicity, and isoelectric point of amino acids encoded allowed for visiting more functional PhoQ variants, the number of functional PhoQ variants was greater for genetic codes allowing for smaller number of synonymous mutations, and for codes with higher fragility and mutability. Such a combination of properties allows for visiting, via point mutations, a greater number of nucleotide sequences while restricting the sequence of PhoQ mutants visited to functional PhoQ mutants, thereby allowing for the exploration of a larger fraction of the space of functional PhoQ variants. Further, code changeability, defined as the sum of probabilities of transitions between all pairs of amino acids, was negatively correlated with the number of functional PhoQ variants explored for type $T_{DNP}$ codes. This result indicates that frequent transitions between different amino acids are not sufficient for exploring a larger fraction of the functional nucleotide sequence space. In fact, such frequent transitions may hamper functional nucleotide sequence space exploration, unless constrained. We have shown that the standard genetic code embodies an evolutionary advantageous balance of robustness and changeability, leading to the exploration of a large fraction of the space of functional protein variants. A similar postulate was previously put forward by Firnberg and Ostermeier in the context of TEM-1 $\beta$-lactamase (Firnberg and Ostermeier 2013).

An organization of codon-amino acid assignments in the genetic code that allows for the exploration of a larger fraction of the space of functional nucleotide sequences via point mutations is unlikely to provide any evolutionary advantage to an individual. Yet, selection for such a property need not invoke a teleological view of evolution. In the vertical descent model of emergence of a universal genetic code (Crick 1968; Wong 1976; Harris et al. 2003), exploration of a larger fraction of the functional nucleotide sequence space will allow the population descended from an individual to access a much larger innovation pool of functional sequences. Thus, such a genetic code will provide a fitness advantage to the population of individuals, particularly amidst changing environmental conditions, and can therefore be selected for. Along similar lines, in the competition between innovation pools model of emergence of a universal genetic code (Vetsigian et al. 2006; Goldenfeld and Woese 2007), a community with a genetic code that allows for the exploration of a larger fraction of the space of functional nucleotide sequences will have access to larger innovation pool. This property will provide a fitness advantage to the individuals in the community, and such a community is more likely to drive out other communities from a niche, allowing for positive selection for the property to explore a larger fraction of the space of functional nucleotide sequences during the emergence of a universal genetic code.

As described above, the property of a genetic code to allow for the exploration of a greater number of functional nucleotide sequences via single nucleotide substitutions affords greater standing diversity in the population. Diversity in the population will result in a fitness advantage under changing environmental conditions. Environment changes are more likely to occur at intermediate time scales as compared to very short or very long time scales. As shown in the Materials and Methods section, each simulation step roughly corresponds to $8.6 \times 10^3/N_e$ years where $N_e$ is the effective population size. For $N_e = 10^6$, 10000 simulation steps will correspond to a period of around 100 years, an approximate time scale for environmental changes. Our results indicate that the standard genetic code is more optimized for exploring a larger fraction of the functional nucleotide sequence space at these intermediate time scales than at very short or very long time scales. Further, the preservation of chemical properties of amino acids under single nucleotide substitutions for which the standard genetic code is well-known to be optimized aids exploration of more functional nucleotide sequences at such intermediate time scales. Thus, the organization of codon-amino acid assignments in the standard genetic code not only helps minimize the phenotypic effects of mutations and translation errors, but also allows for greater standing genetic diversity in the population at the intermediate time scales of typical environment changes. Together, these benefits afforded by the standard genetic code may have contributed towards its emergence as the universal genetic code.

The present study only considers the space of functional variants of the PhoQ protein with amino acid substitutions at 4 positions. Thus, the ideas described above are universal to the extent to which our results for the PhoQ protein can be generalized. Given the dependence of protein function on protein structure and chemistry (Berg et al. 2002), both of which derive from the amino acid composition of the protein, we expect the standard genetic code to exhibit a similar characteristic for functional sequence spaces of other proteins. Comprehensive studies of functionalities of variants of other proteins are needed for strengthening evidence in support of the ideas presented here.

# Materials and Methods

### Generation of type $T_{DP}$ genetic codes

The 64 codons were divided into 21 classes, 20 classes consisting of codons coding for each same amino acid and 1 class consisting of the 3 stop codons. To generate a random code, an amino acid was randomly assigned to one of the 20 classes of codons, not including the class consisting of stop codons. The set of stop codons was left unaltered.

### Generation of type $T_{DNP}$ genetic codes

The set of stop codons was kept the same as in the standard genetic code. Each amino acid was assigned to one codon chosen randomly from among the 61 codons. Each of the remaining 41 codons was then assigned to a randomly chosen amino acid.

### Simulation

In a 12-nucleotide sequence encoding the 4 amino acids at positions 284, 285, 288, and 289 of the *Escherichia coli* protein kinase PhoQ or of the functional variants of this protein (Podgornaia and Laub 2015), one position was chosen randomly, and the nucleotide at that position was mutated to one of the other three possible nucleotides. The mutated sequence was translated into a 4-amino acid sequence using the genetic code being considered, i.e. either the standard genetic code or a randomly generated code. This mutation corresponded to one simulation step. If this new 4-amino acid sequence corresponded to a functional PhoQ variant (Podgornaia and Laub 2015), the mutated nucleotide sequence became the start sequence in the next simulation step. Otherwise, the un-mutated nucleotide sequence remained the start sequence in the subsequent simulation step. Simulations were run using the standard genetic code, 10000 type $T_{DP}$ codes, or 10000 type $T_{DNP}$ codes for 100, 1000, 10000, 100000, or

1000000 steps. With the standard genetic code, the starting 12-nucleotide sequence in the first simulation step was the wild type nucleotide sequence coding for PhoQ in *E. coli*. With a randomly generated genetic code, the starting sequence in the first simulation step was such that it coded for the wild type PhoQ amino acid sequence under the given code, with codons from degenerate sets chosen with probabilities proportional to their frequency in the *E. coli* genome. For each code, for each number of simulation steps, the average number of functional PhoQ variants explored in 100 different simulation runs was reported.

**Definition and calculation of different characteristic measures of genetic codes**

**Mean squared change in physio-chemical properties:** Let $w$ be the amino acid physio-chemical property being considered. Then, the mean squared change in $w$ for a given genetic code $G$ was defined as

$$w_G = \frac{1}{549}\sqrt{\sum (w_{new} - w_{old})^2} \quad (1)$$

where the sum is over all possible single nucleotide mutations, $w_{old}$ is the amino acid encoded by the un-mutated codon, and $w_{new}$ is the amino acid encoded by the mutated codon. The quantity was calculated for the following amino acid properties: polar requirement (C. R. Woese et al. 1966), hydrophilicity (Weber and Lacey 1978), isoelectric point (Alff-Steinberger 1969), and volume (Zamyatnin 1972). References adjacent to each property indicate the study from which the values for the property were taken.

**Code fragility:** Code fragility for a genetic code was defined as the number of codons in the given genetic code for which, out of the 9 possible single nucleotide substitutions, 8 or more were non-synonymous (Judson and Haydon 1999). Note that all genetic codes with the same degeneracy have the same value of code mutability.

**Code mutability:** Let $x_i$ be the number of non-synonymous single nucleotide substitutions for codon $i$ under a given genetic code $G$. Code mutability $M$ for that genetic code was then defined as (Judson and Haydon 1999)

$$M_G = \frac{1}{64}\sum_{i=1}^{64} x_i \quad (2)$$

Note that all genetic codes with the same degeneracy have the same value of code mutability.

**Total number of synonymous point mutations:** Under a genetic code $G$, let $x_i$ be the number of single nucleotide substitutions to codon $i$ that do not change the amino acid encoded by the codon. Then, the total number of synonymous mutations is defined as

$$N_{S,G} = \sum_{i=1}^{64} x_i \quad (3)$$

Note that all genetic codes with the same degeneracy have the same value for this measure.

**Code changeability:** The definition of code changeability was taken from Maeshiro and Kimura (Maeshiro and Kimura 1998). If $i$ and $j$ are amino acids such that it is possible to transition from amino acid $i$ to amino acid $j$ under the genetic code $G$ via a single nucleotide substitution, the probability of transition from $i$ to $j$, $\rho_{ij}$, was defined as

$$\rho_{ij} = \frac{m_{ij}}{9n_i} \quad (4)$$

where $m_{ij}$ is the number of ways of going from a codon that encodes amino acid $i$ to a codon that encodes amino acid $j$ in the genetic code $G$, and $n_i$ is the number of codons that encode amino acid $i$ in the genetic code $G$. For amino acids $i$ and $j$ connected via 2 transitions, $\rho_{ij}$ was defined as

$$\rho_{ij} = \sum_k \frac{m_{ik}}{9n_i} \times \frac{m_{kj}}{9n_k} \quad (5)$$

where $k$ is the intermediate amino acid via which the transition from $i$ to $j$ must take place. The value of $\rho_{ij}$ for amino acids connected via more than 2 transitions was defined along similar lines. The path between amino acids $i$ and $j$ involving minimum number of transitions was chosen for calculating $\rho_{ij}$. The three stop codons were treated as if coding for a 21st amino acid. However, paths that passed through stop codons were excluded from the calculation of $\rho_{ij}$. Code changeability was finally defined as

$$\rho^G = \frac{1}{210} \sum_i \sum_j \rho_{ij}, i \neq j \quad (6)$$

Note that all genetic codes with the same degeneracy have the same value of code changeability.

**Estimate of the time scale corresponding to one simulation step**

Each simulation step corresponds to the time taken for an individual in a population of effective size $N_e$ to acquire one nucleotide substitution in the 12-nucleotide PhoQ sequence considered here. Considering a typical *E. coli* genome size of $5.44 \times 10^6$ base pairs and a mutation rate of 0.003 mutations per genome per generation, one simulation step will be equivalent to

$$\frac{5.44 \times 10^6}{0.003 \times 12 \times N_e} = \frac{1.5 \times 10^8}{N_e} generations \quad (7)$$

With an average generation time of 30 minutes, one simulation step is equivalent to

$$\frac{1.5 \times 10^8}{N_e} \times (30 \times 1.9 \times 10^{-6}) = \frac{8.6 \times 10^3}{N_e} years \quad (8)$$

# Acknowledgements

This work was supported by the Center for Theoretical Biological Physics and funded by the National Science Foundation (PHY-1427654).

# Figures

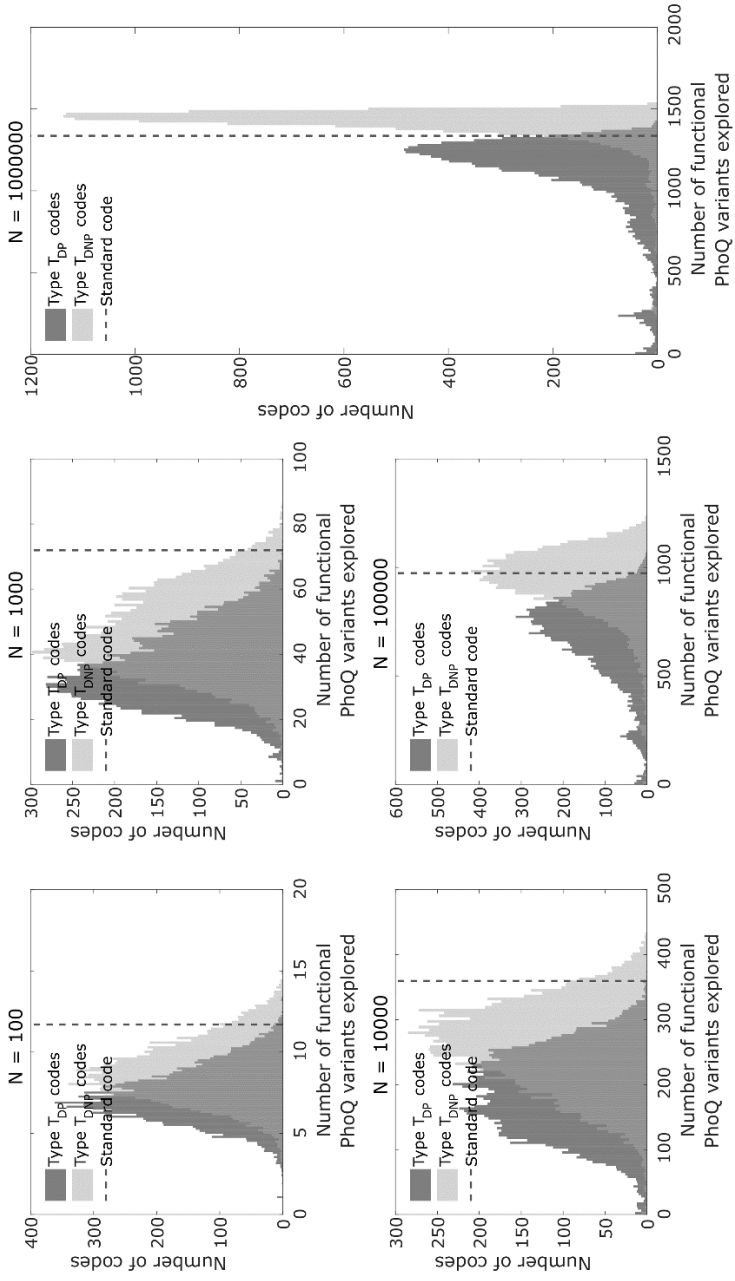

Figure 1

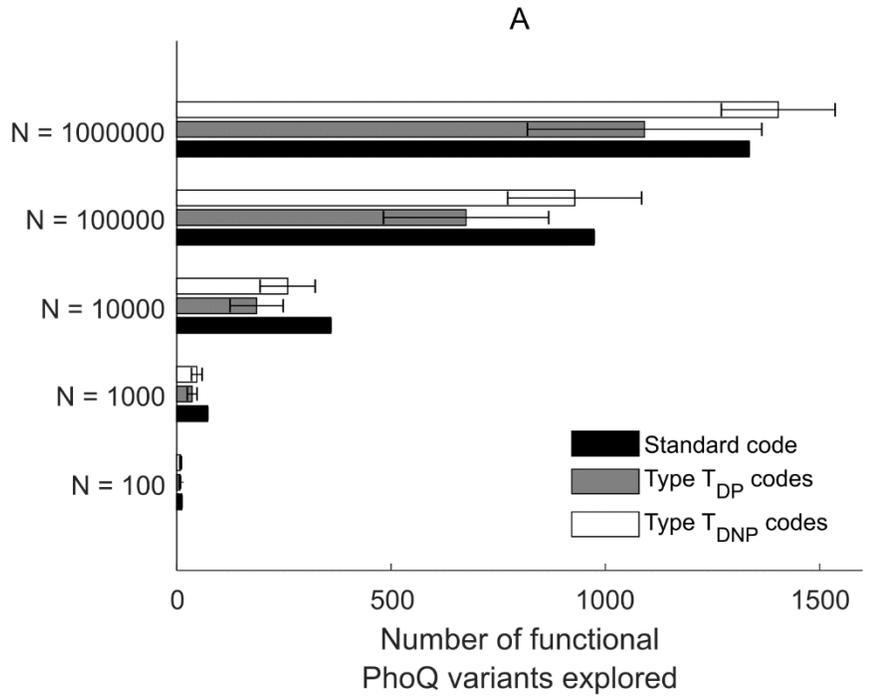

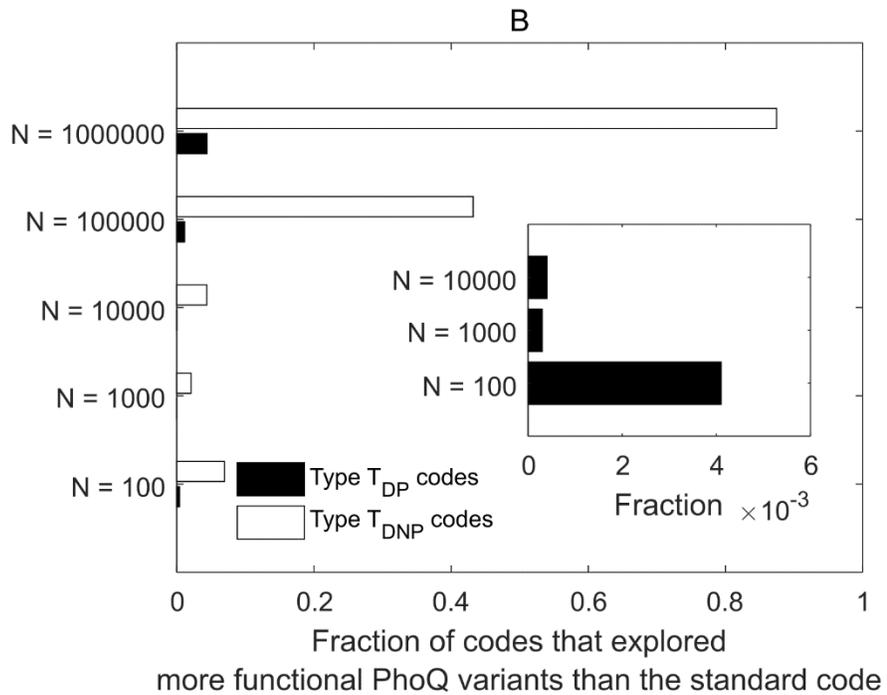

Figure 2

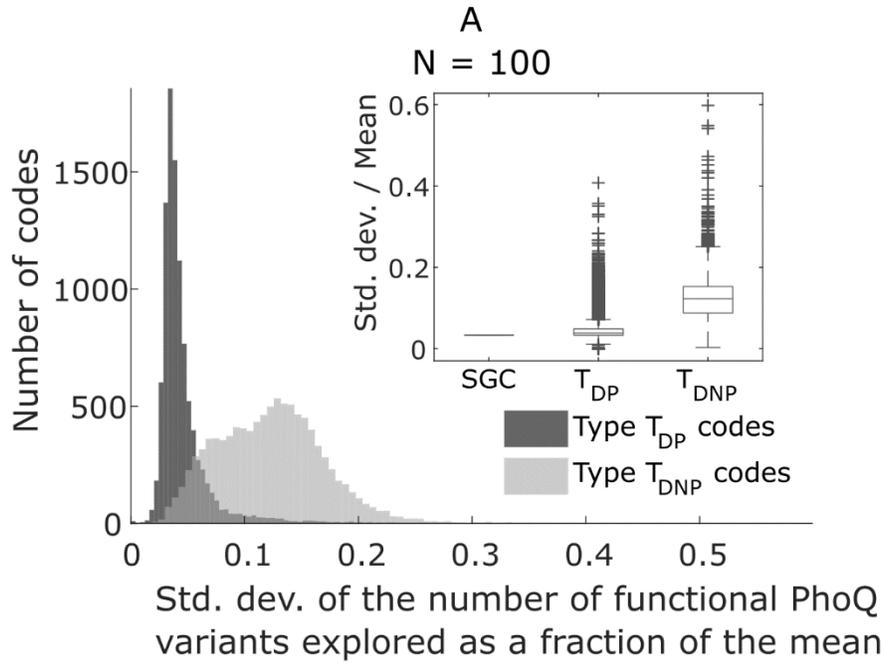
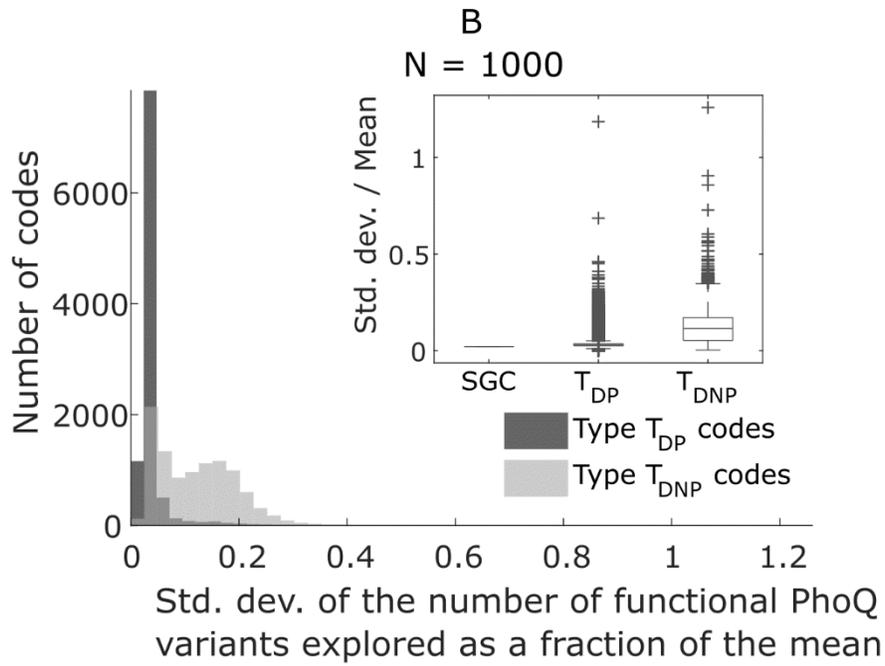

Figure 3

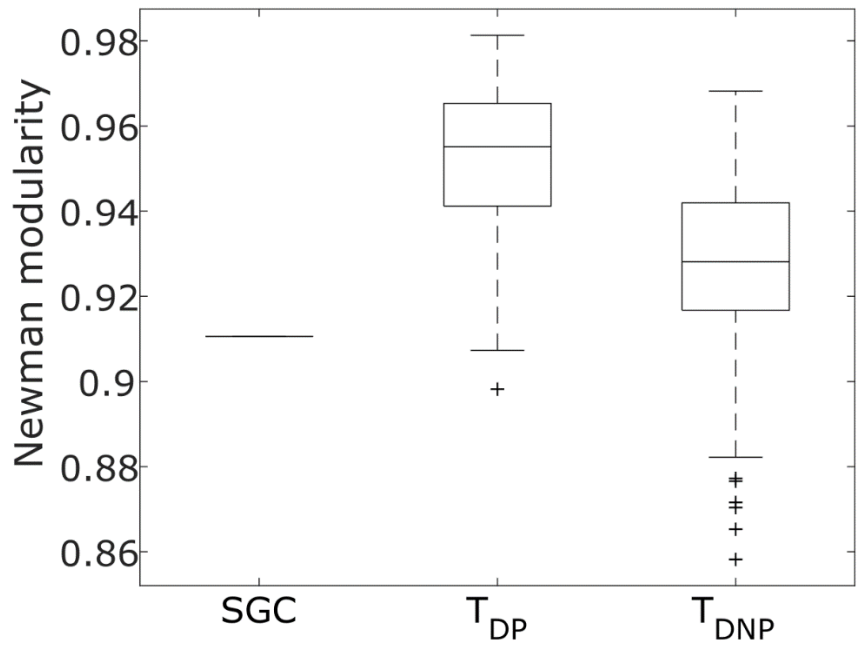

Figure 4

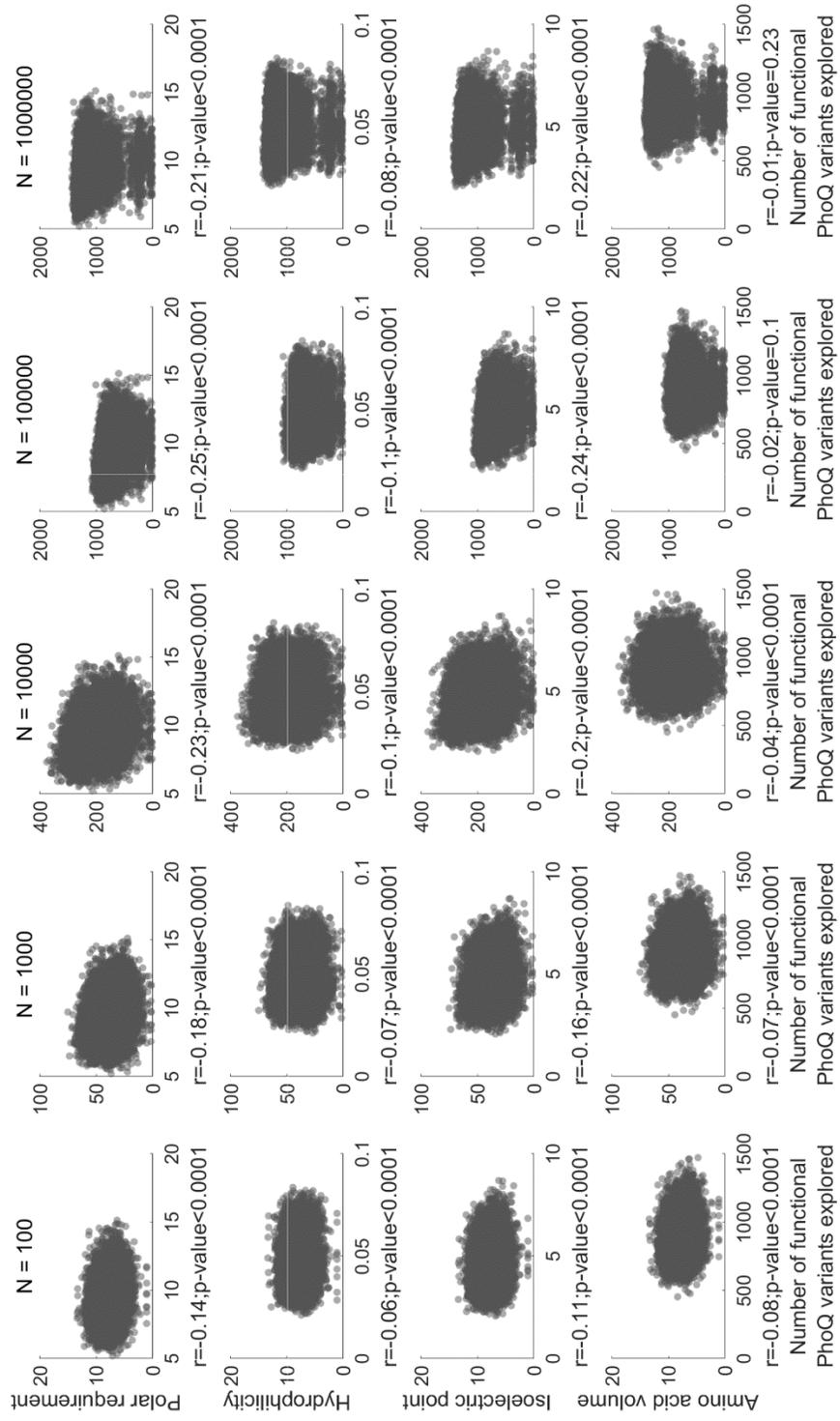

Figure 5

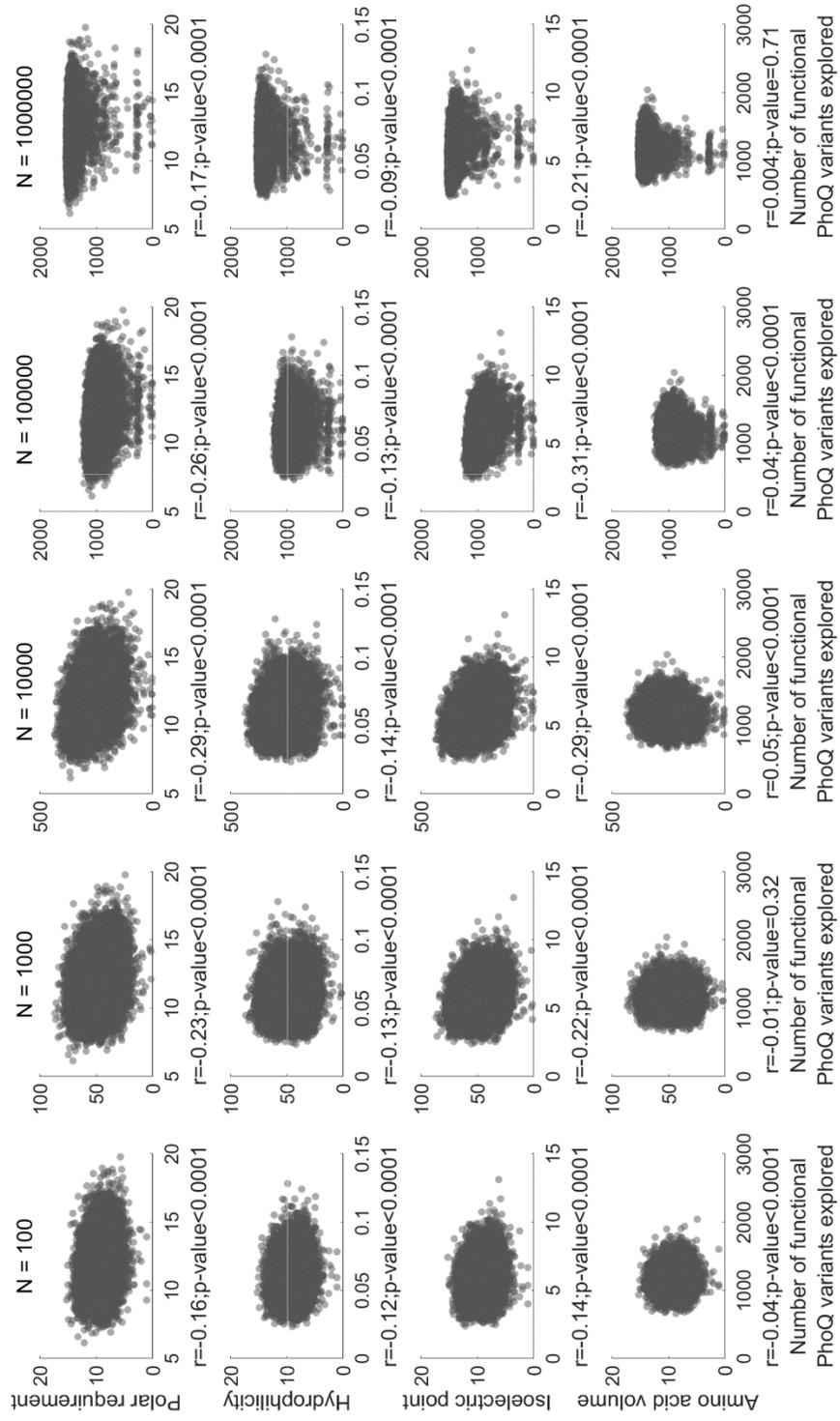

Figure 6

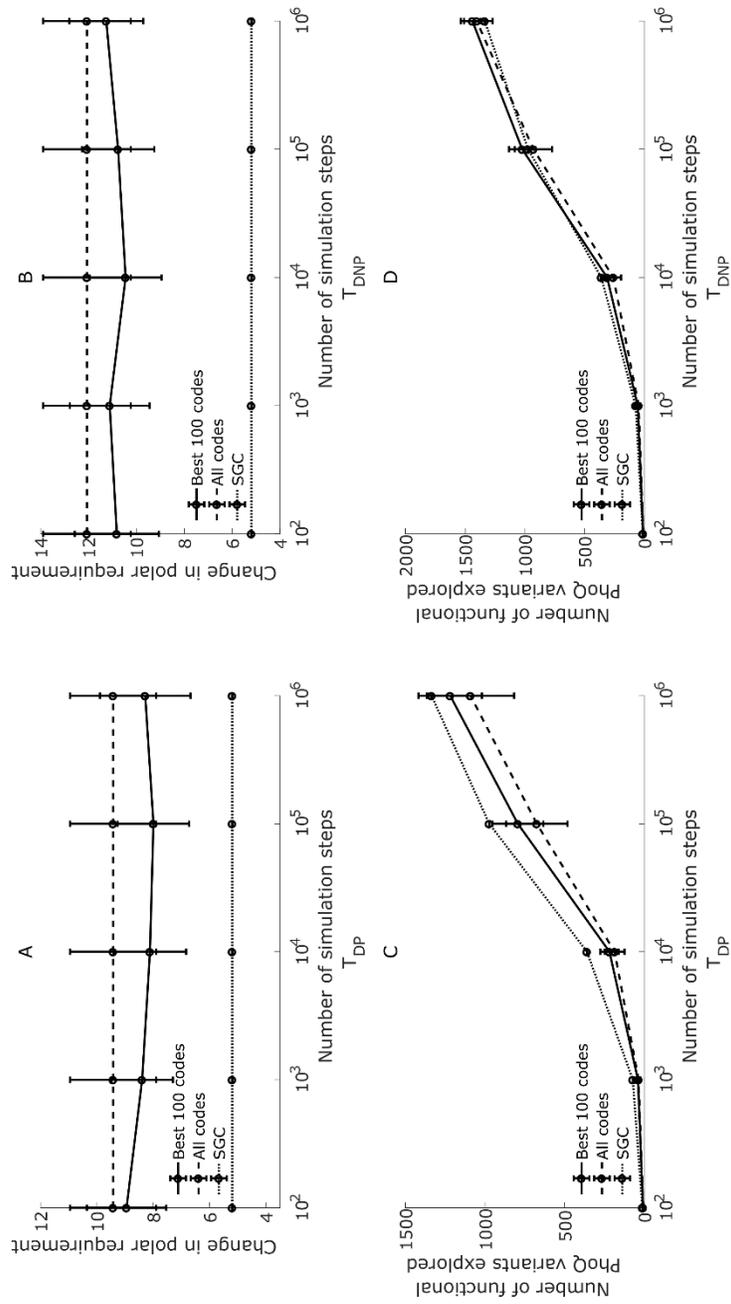

Figure 7

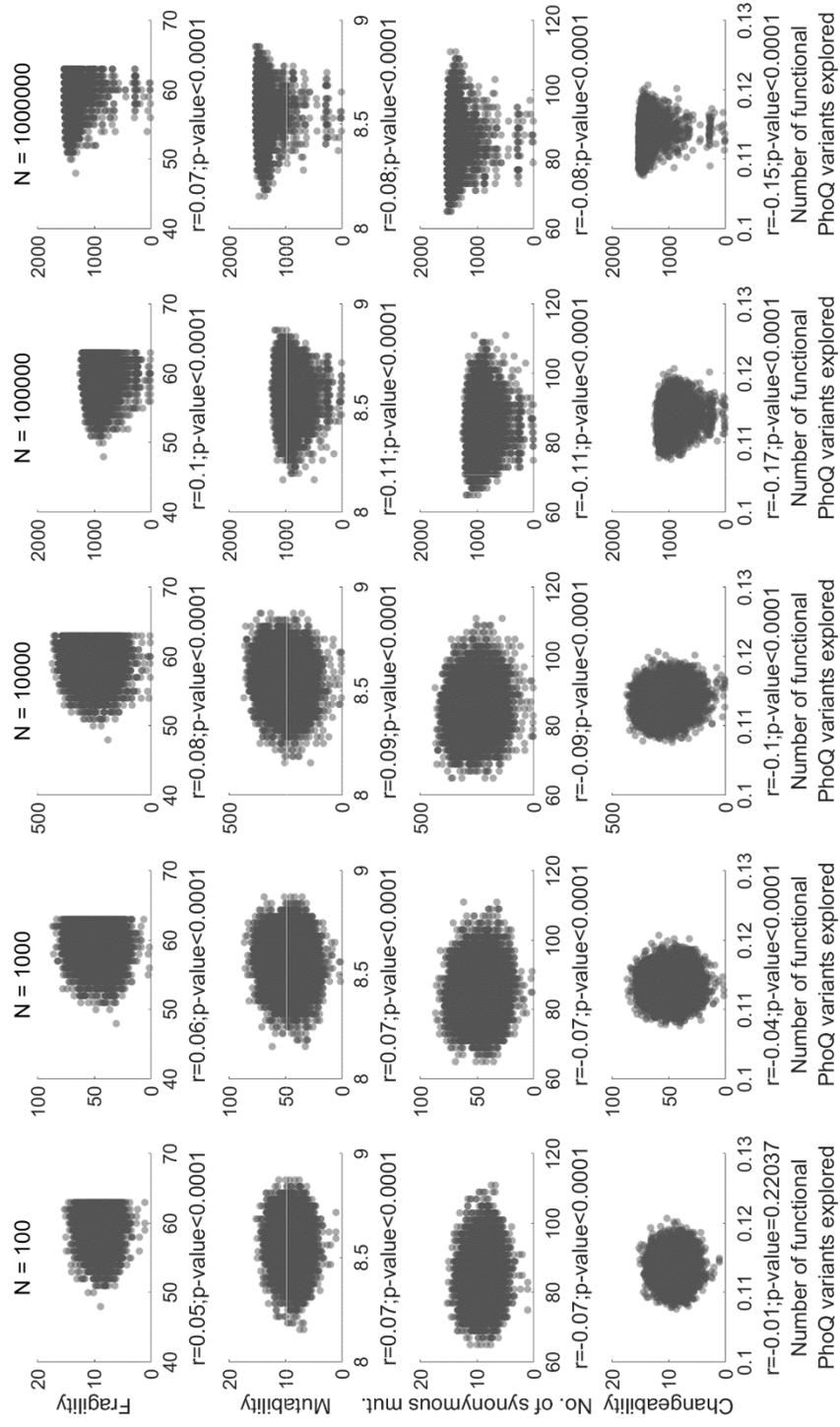

Figure 8

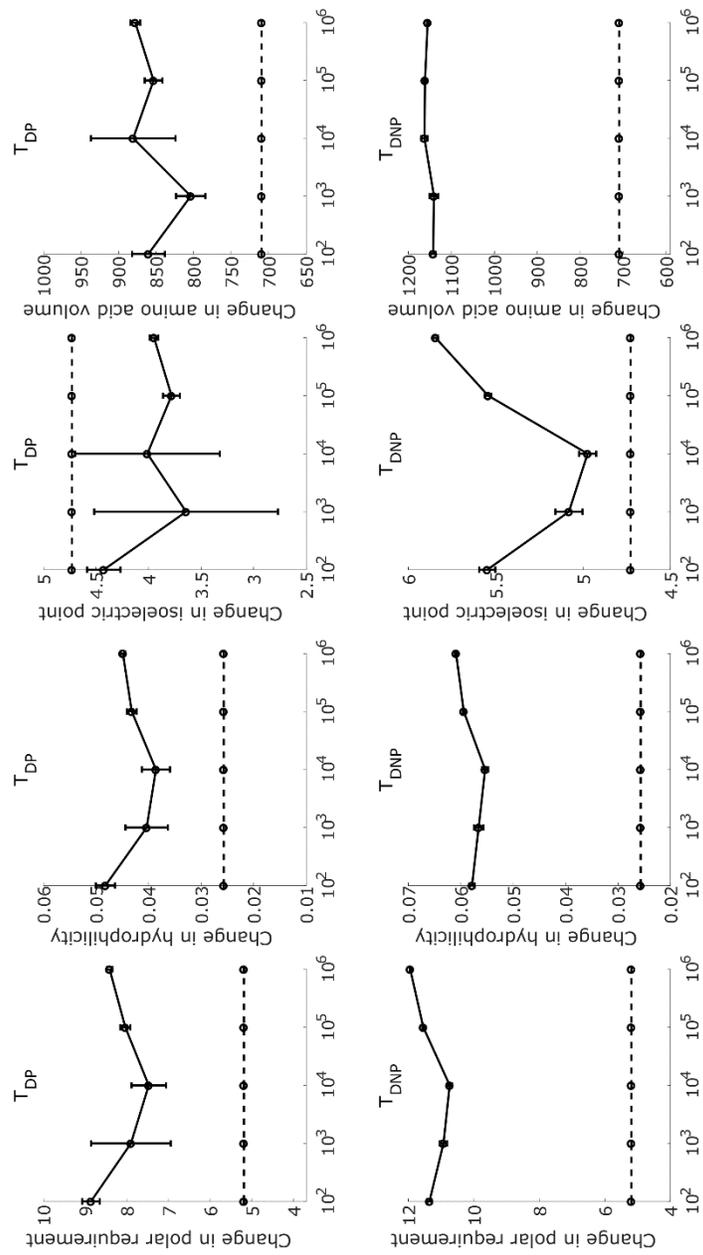

Figure 9

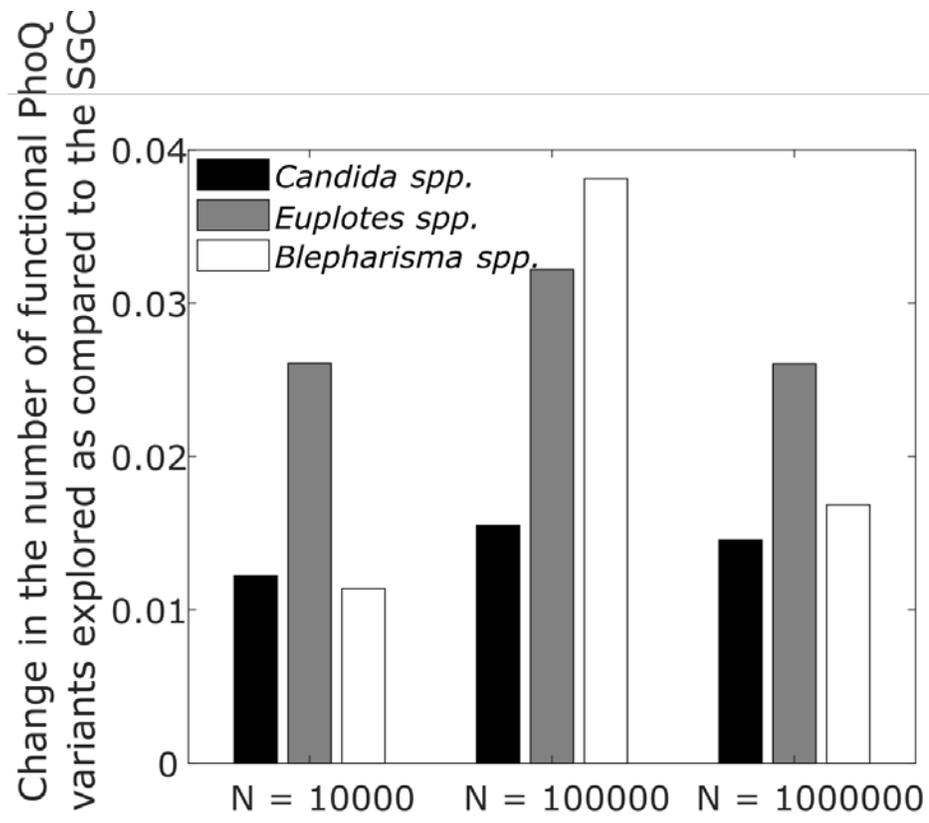

Figure 10

# Figure legends

Figure 1 Distribution of the number of functional variants of the *E. coli* kinase PhoQ explored via single nucleotide substitutions under translation using type $T_{DP}$ codes (pink), type $T_{DNP}$ codes (green), and the standard genetic code (blue line). All simulations were carried out as described in the Materials and Methods section. Distributions are shown for 100, 1000, 10000, 100000, and 1000000 simulation steps.

Figure 2 (A) The number of functional variants of the *E. coli* kinase PhoQ explored via single nucleotide substitutions under translation using different genetic codes for different numbers of simulation steps. The bars indicate the number of functional PhoQ variants explored, averaged over 10000 different type $T_{DP}$ or type $T_{DNP}$ codes. The error bars for randomly generated codes represent one standard deviation from the mean. (B) The fraction of type $T_{DP}$ codes (blue) and the fraction of type $T_{DNP}$ codes (yellow) that allowed for the exploration of more functional PhoQ variants via single nucleotide substitutions for different numbers of simulation steps as compared to the standard code. Inset: Fraction of type $T_{DP}$ codes that allowed for the exploration of more functional PhoQ variants as compared to the standard code for 100, 1000, and 10000 simulation steps.

Figure 3 Distribution of the standard deviation, as a fraction of the mean, of the number of functional PhoQ variants explored on starting the simulation from different degenerate 12-nucleotide sequences that code for the wild type PhoQ amino acid sequence under type $T_{DP}$ and type $T_{DNP}$ codes. The distributions are over 10000 randomly generated codes. (A) Distribution for simulations with $N = 100$ simulation steps. (B) Distribution for simulations with $N = 1000$ simulation steps. Insets: The standard deviation, as a fraction of the mean, of the number of functional PhoQ variants explored on starting the simulation from different degenerate nucleotide sequences for the standard genetic code (SGC), type $T_{DP}$ codes, and type $T_{DNP}$ codes. Inset in (A): Summary statistics for the case of $N = 100$ simulation steps. Inset in (B): Summary statistics for the case of $N = 1000$ simulation steps.

Figure 4 Newman modularity of the network of functional 12-nucleotide sequences coding for the PhoQ protein under the standard genetic code (SGC), 100 type $T_{DP}$ codes, and 100 type $T_{DNP}$ codes. For each genetic code, all 12-nucleotide sequences that translated into a functional 4-amino acid sequence were identified. These sequences formed the nodes of the network. If it was possible to go from one nucleotide sequence to another sequence via a single nucleotide substitution, the nodes corresponding to the two nucleotide sequences were connected via an edge.

Figure 5 Scatter plots representing the dependence of the number of functional PhoQ variants explored under translation using type $T_{DP}$ codes (vertical axis) on the mean squared change in different physio-chemical properties of amino acids due to single nucleotide substitutions: polar requirement, hydrophilicity, isoelectric point, and volume

(horizontal axis). The Pearson's correlation coefficient (r) and the p-value of the estimate are indicated under each plot.

Figure 6 Scatter plots representing the dependence of the number of functional PhoQ variants explored under translation using type $T_{DNP}$ codes (vertical axis) on the mean squared change in different physio-chemical properties of amino acids due to single nucleotide substitutions: polar requirement, hydrophilicity, isoelectric point, and volume (horizontal axis). The Pearson's correlation coefficient (r) and the p-value of the estimate are indicated under each plot.

Figure 7 (A) Mean squared change in polar requirement for the 100 type $T_{DP}$ codes that allowed for the exploration of highest numbers of functional PhoQ variants for different numbers of simulation steps (Best 100 codes) and for all 10000 type $T_{DP}$ codes (All codes). (B) Mean squared change in polar requirement for the 100 type $T_{DNP}$ codes that allowed for the exploration of highest numbers of functional PhoQ variants for different numbers of simulation steps (Best 100 codes) and for all 10000 type $T_{DNP}$ codes (All codes). (C) Number of functional PhoQ variants explored under translation for different numbers of simulation steps using the 100 type $T_{DP}$ codes with least mean squared changes in polar requirement (Best 100 codes), 10000 type $T_{DP}$ codes (All codes), and the standard genetic code (SGC). (D) Number of functional PhoQ variants explored under translation for different numbers of simulation steps using the 100 type $T_{DNP}$ codes with least mean squared changes in polar requirement (Best 100 codes), 10000 type $T_{DNP}$ codes (All codes), and the standard genetic code (SGC). The error bars represent one standard deviation from the mean.

Figure 8 Scatter plots representing the dependence of the number of functional PhoQ variants explored under translation using type $T_{DNP}$ codes (vertical axis) on quantitative measures characterizing the organization in codon-amino acid assignments: code fragility, code mutability, total number of synonymous point mutations, and code changeability (horizontal axis). Detailed definitions of these properties are given in the Materials and Methods section. The Pearson's correlation coefficient (r) and the p-value of the estimate are indicated under each plot.

Figure 9 Mean squared changes in different physio-chemical properties of amino acids due to single nucleotide substitutions for those type $T_{DP}$ codes (top row; blue) and type $T_{DNP}$ codes (bottom row; red) that explored more functional PhoQ variants as compared to the standard code (green) for 100, 1000, 10000, 100000, and 1000000 simulation steps. Results are shown for amino acid polar requirement, hydrophilicity, isoelectric point, and amino acid volume. The error bars represent one standard error of the mean.

Figure 10 Fraction change in the number of functional PhoQ variants explored under translation using deviant genetic codes from different species as compared to the standard genetic code (SGC) for number of simulation steps $N = 10000$, $N = 100000$, and $N = 1000000$. The change is calculated as $\Delta f = (f_{code} - f_{SGC})/f_{SGC}$. Here, $f_{SGC}$ is the mean of the number of functional PhoQ variants explored using the standard

genetic code over 100 different simulations and $f_{code}$ is the mean of the number of functional PhoQ variants explored using the deviant genetic code over 100 different simulations.

# Tables

Table 1 p-values for the number of functional PhoQ variants explored under translation using deviant genetic codes from different species compared to the number explored under translation using the standard genetic code.

| Species | Codon re-assignment | $N=10^2$ | $N=10^3$ | $N=10^4$ | $N=10^5$ | $N=10^6$ |
|---|---|---|---|---|---|---|
| *Candida* spp. | CTG; Leu → Ser | 0.0158 | 0.0466 | <0.01 | <0.01 | <0.01 |
| *Mycoplasma* spp. | TGA; Stp → Trp | 0.3822 | 0.9664 | 0.9533 | <0.01 | <0.01 |
| *Euplotes* spp. | TGA; Stp → Cys | 0.3664 | 0.1510 | <0.01 | <0.01 | <0.01 |
| *Blepharisma* spp. | TAG; Stp → Gln | 0.5101 | 0.0267 | <0.01 | <0.01 | <0.01 |